\documentclass[floats,floatfix,showpacs,amssymb,amsmath,prl,twocolumn,superscriptaddress,nofootinbib, aps]{revtex4-2}

\usepackage{dcolumn}
\usepackage{bm}
\usepackage{aas_macros}
\usepackage{graphicx}	
\usepackage{amssymb, amsmath}
\usepackage{natbib}
\usepackage{hyperref}
\hypersetup{colorlinks,bookmarksnumbered,
citecolor=cyan, urlcolor=magenta}

\providecommand{\adsurl}[1]{\href{#1}{ADS}}
\usepackage[usenames, dvipsnames]{xcolor}

\begin{document}
\preprint{APS/123-QED}

\title{What is the price of abandoning dark matter?\\Cosmological constraints on alternative gravity theories}
\author{Kris Pardo}
\email{kpardo@caltech.edu}
\affiliation{Jet Propulsion Laboratory, California Institute of Technology, Pasadena, CA 91101, USA}
\affiliation{Department of Astrophysical Sciences, Princeton University, Princeton NJ 08544, USA}
\author{David N. Spergel}
\affiliation{Department of Astrophysical Sciences, Princeton University, Princeton NJ 08544, USA}
\affiliation{Center for Computational Astrophysics, Flatiron Institute, NY NY 10010, USA}
\date{\today}
\begin{abstract}
Any successful alternative gravity theory that obviates the need for dark matter must fit our cosmological observations. Measurements of microwave background polarization trace the large-scale baryon velocity field at recombination and show very strong, $O(1)$, baryon acoustic oscillations. Measurements of the large-scale structure of galaxies at low redshift show much weaker features in the spectrum. If the alternative gravity theory's dynamical equations for the growth rate of structure are linear, then the density field growth can be described by a Green's function: $\delta(\vec x,t) = \delta(\vec x,t')G(x,t,t')$. We show that the Green function, $G(x,t,t')$, must have dramatic features that erase the initial baryon oscillations. This implies an acceleration law that changes sign on the $\sim 150$ Mpc scale. On the other hand, if the alternative gravity theory has a large nonlinear term that couples modes on different scales, then the theory would predict large-scale non-Gaussian features in large-scale structure. These are not seen in the distribution of galaxies nor in the distribution of quasars. No proposed alternative gravity theory for dark matter seems to satisfy these constraints.
\end{abstract}
\maketitle

\section{Introduction}

The astronomical evidence for dark matter continues to grow: the velocities of galaxies imply the existence of dark matter in clusters \citep{Zwicky1933,Einasto1974}; measurements of rotation curves reveal its presence in galaxies like our own \citep{Rubin1978,BosmaPhD,Rubin1980}; dynamical arguments demonstrate its ubiquity \citep{Ostriker1974}; and gravitational lensing measurements confirm its presence in clusters and galaxies \citep{Tyson1990,Wittman2000}. Cosmological observations provide another line of evidence for the existence of dark matter: the popular $\Lambda$ Cold Dark Matter ($\Lambda$CDM) model is remarkably successful in simultaneously fitting cosmic microwave background (CMB) observations and the large-scale distribution of structure \citep[e.g.,][]{Spergel2003, Planck2018}. This concordance requires that the dominant form of matter is not baryons but cold, weakly-interacting (or non-interacting) dark matter.

While dark matter has become part of the standard paradigm, we have yet to detect it. With ever improving dark matter experiments ruling out much of the parameter space associated with the ``WIMP" miracle \citep{Aprile2018,aprile2019}, there has been renewed interest in alternative gravity theories that obviate the need for dark matter.

However, it has proven very challenging to develop a satisfactory alternative to General Relativity (GR). Any successful modified gravity theory will need to reproduce the successes of $\Lambda$CDM and GR:
\begin{enumerate}
\item Provide an explanation for the flatness of galaxy rotation curves at large radii, the distribution of hot gas in elliptical galaxies and clusters of galaxies, and match the gravitational lensing shear measurements;
\item Satisfy the classical tests of GR, including the precession of the perihelion of Mercury and other solar system tests, the Shapiro time delay, and the timing of binary millisecond pulsars \citep{Will1993}.
\item Provide a consistent fit to LIGO's gravitational wave signals. These measurements provide strong constraints on the tensor content of any gravitational wave theory \citep{LIGOGRTests2019,Abbott2017,Pardo2018, Lagos2019, Boran2018}.
\item Predict an expanding universe and provide an acceptable fit to measurements of the distance-redshift relationship. This constrains the homogeneous cosmological solution of the alternative theory \citep[c.f.,][]{Ishak2019}.
\item Provide a satisfactory fit to measurements of both the CMB fluctuations and the large-scale structure. 
\end{enumerate}

This Letter quantifies the final constraint on this list: any alternative gravity theory that obviates the need for dark matter needs to provide an explanation for the growth and evolution of structure. 

The $\Lambda$CDM model accurately explains how structure forms from initial density perturbations and how these perturbations are imprinted in the cosmic microwave background \citep{Lifshitz1946, Peebles1970, Sunyaev1970b,Bond1984, Peebles1984}. The initial fluctuations are adiabatic: overdense regions have an excess of baryons, dark matter, and photons. In the early universe, the fluctuations in the tightly-coupled baryon-photon fluid oscillate like sound waves. On the other hand, the dark matter is cold and its fluctuations evolve only through gravity. After recombination, baryons decouple from the photons and then fall into the growing dark matter potential wells. This dark matter driven gravitational fluctuation growth erases most of the signature of the sound waves. Thus, the $\Lambda$CDM model can explain why the acoustic oscillations in the cosmic microwave background temperature and polarization fluctuations have $O(1)$ amplitude and the oscillations in the distribution of galaxies are subtle with amplitude of $O\left((\Omega_b/\Omega_m)^2\right) \sim 0.04$. {\it Any alternative gravity theory will have to provide an alternative explanation for this suppression of the acoustic fluctuations, one of the distinctive effects of dark matter}.

In this Letter, we outline how to determine the required infrared (IR) behavior of any dark matter theory based on linking the baryon density field at recombination ($z\sim1100$) to the baryon power spectrum at low redshift ($z\sim 0$). Any successful theory for dark matter, whether it invokes particles or alternative theories of gravity, must properly explain this evolution. These density fields are typically probed indirectly through fitting the CMB power spectra and the matter power spectrum in tandem \citep[e.g.,][]{Spergel2003, Planck2018}. This necessarily assumes $\Lambda$CDM (or some simple extension), as well as GR. The test we propose here does not invoke GR nor a specific cosmology. Instead it relies solely on small-scale physics -- Thomson scattering and the Newtonian continuity equation. Note that while similar tests have been proposed before \citep{McGaugh2004, Dodelson2011}, they have not been explicitly formulated nor calculated for general modified gravity theories.

The polarization of the CMB on small scales is exclusively due to Thomson scattering, which itself only relies on the velocities of the electrons. Because protons and electrons are tightly coupled via Coulomb scattering at early times, we can assume that the velocities of the electrons exactly equals that of the protons. The CMB polarization spectrum then directly measures the velocity of the baryons at $z\sim1100$. The Newtonian continuity equation, which is valid at small scales, relates the velocities of the baryons to their density field. Thus, the CMB polarization spectrum is a direct measurement of the baryon velocity field at $z\sim 1100$. At $z\sim 0$, the galaxy-galaxy correlation function traces the baryon density field at large scales. With these two direct measures of the baryon density field, we can then define the form a linear alternate theory of dark matter must take in the IR. We combine observations of the CMB and the galaxy power spectrum at low-redshift to determine the required Green's function of structure formation between these redshifts for alternate theories. This Green's function has a distinctive form as it must suppress the baryon acoustic oscillations by nearly an order of magnitude, as well as greatly increase power on small scales. 

Below we describe the theoretical framework for determining the IR behavior of modified gravity theories for dark matter. We first outline the general idea behind our method, which will depend on the baryon power spectrum at both $z\sim 0$ and $z\sim1100$. We then describe how we calculate each of these power spectra. Finally, we give the resulting necessary form for an alternative dark matter theory and conclude.

\section{Infrared Behavior of Modified Gravity}
We assume that the modified gravity theory predicts our universe is expanding with a scale factor, $R(t)$, determined by its dynamical equations and that the form of $R(t)$ fits the current measurements of the distance-redshift relation. This assumption already places a very profound constraint on any alternative to GR.

As is usual in cosmology, we represent the density field as the sum of a mean density field, $\rho(t)$, and spatial fluctuation: $\rho(\vec x,t) = \rho(t)\left[1 + \delta(\vec x,t)\right]$ and expand the density field in Fourier modes: $\delta(\vec k,t)$, where $\vec{k}$ is an angular wavevector. The power spectrum, $P(k)$, is then given by the spatial, two-point correlation function of these Fourier modes at any one time: $\langle \delta(\vec{k},t) \delta(\vec{k}',t) \rangle = (2\pi)^3 \delta^3(\vec{k}-\vec{k}') P(k)$, where $\delta^3(x)$ is the 3D Dirac delta function.

In alternative gravity theories, the acceleration encodes the deviation from GR -- these theories generally assume matter and momentum conservation. Thus, we will also assume these conservation laws hold. In agreement with the cosmological principle and observations of large scale structure, we will also assume that any modifications to GR must be isotropic. 


We assume that the acceleration in the modified gravity theory only depends on the amplitude of the baryon density fluctuations: $\vec a(\delta_b)$. We then expand the function as a series of sums of Fourier modes:
\begin{equation}
    a(k,t) = \hat{F}_1(k) \delta_b(k,t) + \sum_{k'} \hat{F}_2(k, k') \delta_b(k,t) \delta_b(k',t) + \ldots
\end{equation}
where $k \equiv \lVert\vec{k}\rVert$, $\hat{F}_1(k)$ is the linear response to the density fluctuation (including both GR and modified terms), and $\hat{F}_2$ encodes the second order correction.


Since the density field is small, the linear term should dominate the gravitational acceleration in most modified gravity theories. Thus, we focus on linear modifications to GR in this paper. Note that this linear term acts like a transfer function -- it has no explicit time dependence and is simply multiplied with a given density configuration in $k$-space to give the resulting acceleration force.

If the modified gravity theory has strong nonlinear terms, then the theory will produce significant mode-mode couplings that would be apparent in the large-scale structure. The theory could evade the current strong constraints from Planck on non-Gaussianity \citep{PlanckIX} if the theory is linear at early times. However, if the theory is nonlinear enough at late times to erase the baryon acoustic oscillations, then these same nonlinearities would induce large non-Gaussian features in the large-scale distribution of structure. These are not seen in the large-scale distribution of structure \citep{Slosar2008}. Thus, it is unlikely that a strongly nonlinear theory could produce the correct evolution for the baryons and evade low-redshift non-Gaussianity constraints. Detailed calculations showing this point are left to future work.

\section{Linear Modification to General Relativity\label{sec:linear}}


In $\Lambda$CDM after recombination, baryons fall into the dark matter potentials. This imprints the large-scale distribution of the dark matter on the baryons. Thus, the transfer function of CDM, along with the initial spectrum of fluctuations, is all that is needed to accurately describe the matter power spectrum. The baryon power spectrum follows directly by using the CDM potential created by the evolution of these perturbations. However, if we no longer have CDM in our model, the baryon transfer function itself must encode all of this information. \textit{In modified gravity theories of dark matter, the baryon transfer function must account for all of the changes in the baryon perturbations from early to late times.}

The matter power spectrum depends on the transfer function as: $P(k) \propto P_{\phi}(k)T^2(k)$, where $P_{\phi}$ is the primordial spectrum of perturbations. In analogy to this, we can define the transfer function:
\begin{equation}
    \hat{T}_b^2(k) = \frac{P_{bb}(k, z\sim 0)}{P_{bb}(k,z=1100)} \; .
\end{equation}
$\hat{T}_b^2(k)$ describes how the baryon perturbations evolve from $z=1100$ to $z\sim 0$, where the hat indicates a different normalization than is typically used.

Any theory for dark matter must adequately explain both the shape and normalization of $\hat{T}_b^2(k)$. Our transfer function can be exactly represented with measurable data and does not rely on any assumptions about underlying theories, outside of the small-scale physics described below. It is also possible to find the theoretical solutions for any dark matter or modified gravity theories. In this paper, we will focus solely on the shape of $\hat{T}_b^2(k)$ -- a more precise analysis is required to use the normalization as well.

As a way of building intuition, we will also consider the Fourier pair of the transfer function -- the Green's function:
\begin{equation}\label{eqn:greens}
    \hat{\mathcal{G}}_b(r) = \int dk\ \frac{k^2}{2\pi^2} \hat{T}_b(k) j_0(kr) \; ,
\end{equation}
where $j_0(x)$ is a Spherical Bessel function. This function shows, in real space, the inherent acceleration response of the modified gravity. 


\subsection{The Baryon Power Spectrum at $z\sim0$}\label{sec:theory-lowz}


The baryons at low redshift and large scales ($\gtrsim 10~\rm{Mpc}$) are well-traced by galaxies. Thus, we can take the 3D power spectrum of galaxies as the baryon power spectrum.
We use the data from Ref.~\cite{Beutler2016} for the galaxy-galaxy power spectrum at low-z. Ref.~\cite{Beutler2016} measures the BAO signal from galaxies from $z=0.2-0.75$ using the Sloan Digital Sky Survey-III \citep[SDSS-III;][]{Eisenstein2011} Baryon Oscillation Spectroscopic Survey (BOSS) DR12 data set \citep{Dawson2013,Alam2015}. As part of this measurement, they also calculate the 3D galaxy-galaxy power spectrum in 3 different redshift bins. We use the lowest redshift bin, $z=0.2-0.5$, which has an effective redshift of $z=0.38$. This is measured from $k=0.016 - 0.15~h~\rm{Mpc}^{-1}$. We use their fiducial value of $h=0.676$ to transform to physical units.



\begin{figure}[!htb]
    \centering
    \includegraphics[width=.48\textwidth]{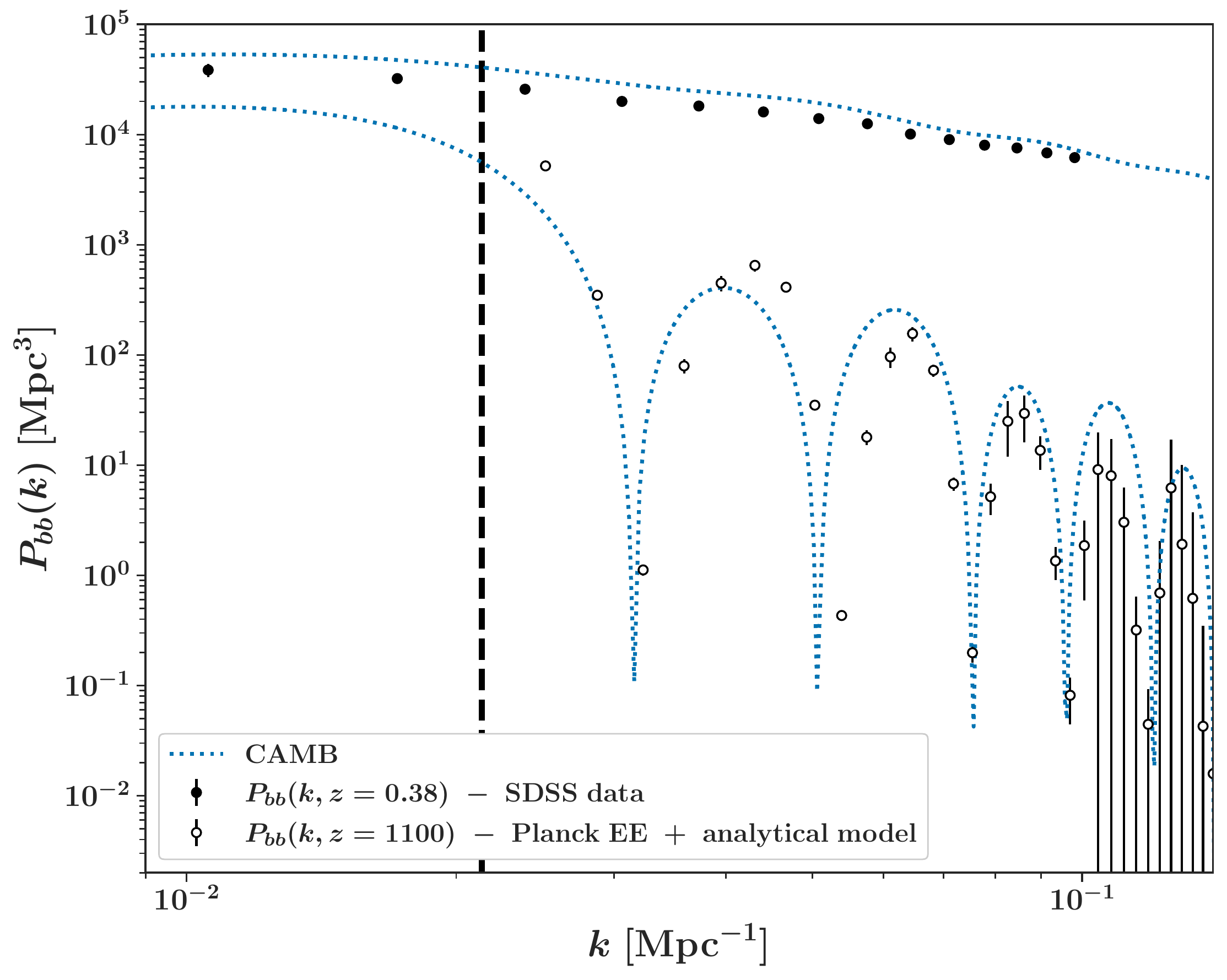}
    \caption{\label{fig:ps} Baryon power spectra at $z=0.38$ (filled-in circles) and $z=1100$ (open-circles). The $z=0.38$ data is taken from Ref.~\cite{Beutler2016}. The $z=1100$ points use the polarization data from Refs.~\cite{Planck2019, ACT2016} and our analytical model, described in the text. The difference between the circles and the spectrum produced by a full treatment by the CAMB code for each redshift (blue dotted line), assuming the Ref.~\cite{Planck2018} values, provides an estimate of the error in the approximation we use. The majority of this error is due to the lack of a DM potential driving term in Equation~\ref{eqn:continuity}. This would shift the peaks to align with the CAMB case. The black, dashed line gives the acoustic scale, as given by Ref.~\cite{Planck2018}. All high-redshift curves are arbitrarily normalized.}
\end{figure}

\subsection{Power Spectrum at ${z\sim1100}$}

The polarization of the CMB can be related to the velocity of the baryons as \citep{Zaldarriaga1997}:
\begin{equation}\label{eqn:QU}
\Delta_p(\hat n, \vec{x}) = Q(\hat n) +iU(\hat n) \approx 0.17 \Delta \tau_* \hat m^i \hat m^{j} \partial_i v_{j}
\end{equation}
where $\Delta_p$ is the polarization fluctuation, $Q$ and $U$ are Stokes parameters, $\hat{n}$ is the direction of observation (i.e. into the sky), $\Delta \tau_{*}$ is the width of the last scattering surface, $\hat{m}$ is a 2D unit vector on the plane of the sky, and $v$ is the baryon velocity on the sky. 




The velocity due to density perturbations is \textit{irrotational}. Thus, the polarization spectrum is just the gradient of the baryon velocity field: $\Delta_p(\hat{n}, \vec{k}) \approx 0.17 \Delta \tau_* ik v_b$. 



Typically, polarization results are reported using $E$ and $B$-modes, which are just a rotation of the $Q$-$U$ basis that sets $B=0$ on small scales in the early Universe. Thus, the polarization power spectrum is just the E-mode power spectrum:
\begin{equation}\label{eqn:EEspec}
    P_{EE}(k) \approx (0.17 \Delta \tau_*)^2 k^2v^2_b(k) \; .
\end{equation}

Prior to recombination, the baryons and photons can be treated as a single fluid. In a universe with no DM, the behavior is simple inside the horizon:
\begin{equation}\label{eqn:continuity}
 \ddot{\delta}_b + c_s^2 k^2 \delta_b = 0  \; ,
\end{equation}
where $\dot{}~ \equiv \frac{d}{d\tau}$ (conformal time) and $c_s$ is the sound speed. 


For adiabatic initial conditions, this admits the solution:
\begin{equation}\label{eqn:deltaz1100}
    \delta_b = A(k) \cos (kr_s) \; ,
\end{equation}
where $r_s = \int d\eta~c_s $ is the sound horizon.


The density can be related to the velocity via the continuity equation. At small scales, we can ignore any changes in the potential and simply treat the baryon-photon fluid as a normal Newtonian fluid. Using the continuity equation in Fourier space, we find:
\begin{equation}
    v_b = \frac{i}{k} \dot{\delta}_b(k) = - i  c_s A(k) \sin(k r_s) \; .
\end{equation}
From Equation~\ref{eqn:EEspec}, we have:
\begin{equation}\label{eqn:EEandvk}
    P_{EE}(k) \approx (0.17 \Delta \tau_*)^2 c_s^2 k^2 |A(k)|^2 \sin^2(kr_s)
\end{equation}
We can find $A(k)$ using the observed $EE$ power spectrum and then use Equation~\ref{eqn:deltaz1100} to find the density power spectrum. Note that velocity overshoot may shift the peak positions here, but will not change the overall shape of the power spectrum. There is also a small effect from the finite thickness of the last scattering surface -- this amplifies scales that are smaller than the thickness of the surface. To account for this effect, we multiply Equation~\ref{eqn:EEandvk} by an exponential factor, $\exp[k/k_{\Delta \tau_*}]$, with $\Delta \tau_* = 19~\rm{Mpc}$~\citep{Hadzhiyska2019}.

For the $EE$ power spectrum, we use the Planck 2018 \citep{Planck2019} and the Atacama Cosmology Telescope ACTPol Two Season \citep{ACT2016} angular power spectra. The data is given as multipoles, $C_l^{EE}$, of the 2D power spectrum, which we convert to the 3D power spectrum, $P_{EE}(k)$, using $l = k\eta_* - \frac{1}{2}$, where $\eta_*$ is the conformal distance to the last scattering surface\footnote{This does require setting a cosmology. We use the measured distance to the last scattering surface from Ref.~\citep{Planck2018}. Since we require that the modified gravity must also fit the measured distance-redshift relation, the distance from last scattering cannot deviate too wildly from the Planck value. In principle, it may be possible to set $\eta_*$ without setting a cosmology -- instead, we might be able to use the alignment of the peaks in each of the power spectra.}\citep{Loverde2008}. Then, to order unity, the 3D power spectrum is \citep{Bond1987, Loverde2008}: $P_{EE}(k) \sim \frac{\pi l^2}{k^3}C_{l=k\eta_{*}-\frac{1}{2}}^{EE}$. We bin the $C_{l}^{EE}$ data into $l$-bins with width $\Delta l = 50$ to increase the signal-to-noise. We also only use $l\leq2000$, due to the high noise in the data above this point.

In Figure~\ref{fig:ps}, we show the baryon power spectrum at $z=1100$ and $z=0.38$. As can be seen, the proper dark matter theory must somehow explain how the $z=1100$ spectrum smooths out and increases in power on small scales. Note that our peaks do not precisely correspond with the CAMB\footnote{\url{https://camb.info/}}-derived peaks at low-$k$. This occurs because we ignore the cold dark matter driving-term in the continuity equation, which is more prominent at low-$k$ (i.e. velocity overshoot; cf. \citep{Sunyaev1970b, Press1980, Hu1995}). 




\begin{figure}[!htb]
    \centering
    \includegraphics[width=0.5\textwidth]{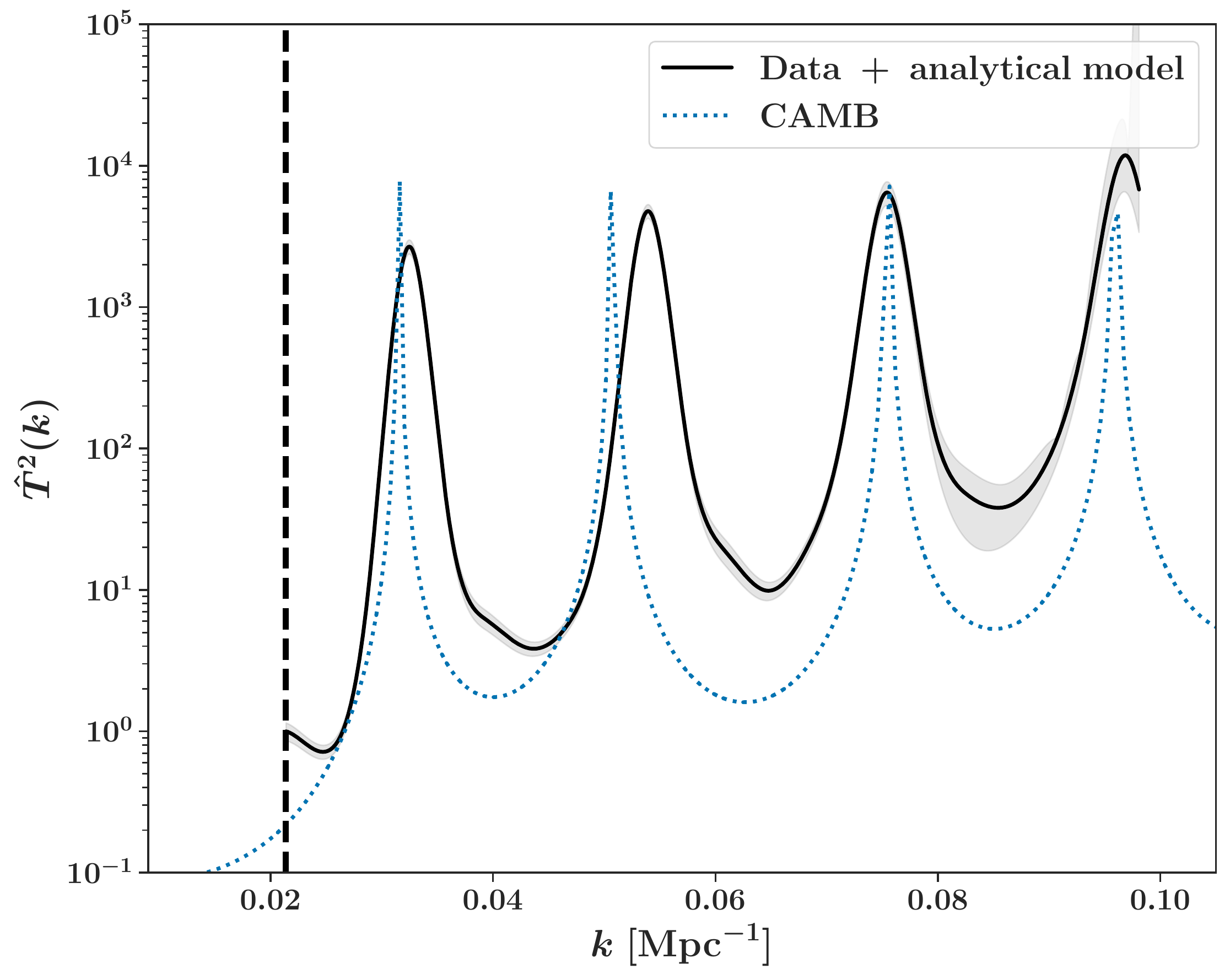}
    \caption{Baryon transfer function from $z=1100$ to $z=0.38$ using the data (black line) and CAMB, assuming $\Lambda$CDM and the values from Ref.~\citep{Planck2018} (blue, dotted line). The difference between the two shows the limitations of the analytical approximation used to derive the transfer function. The gray region shows the 1-$\sigma$ error from the data.}
    \label{fig:transfer}
\end{figure}
\subsection{Constraining linear modified gravity theories}
To derive the transfer and Green's functions for the modified gravity theory we only use the data from each survey where they both overlap in $k$.  


The transfer function is shown in Figure~\ref{fig:transfer}. We also include the CAMB-derived transfer function, which we derive by taking the baryon power spectrum at the same redshifts as our data and dividing them. The transfer function makes the exact evolution of perturbations needed apparent. Power should grow the most on small scales and should oscillate to smooth out the baryon acoustic oscillations. This aligns with the standard $\Lambda$CDM picture.

\begin{figure}
    \centering
    \includegraphics[width=0.5\textwidth]{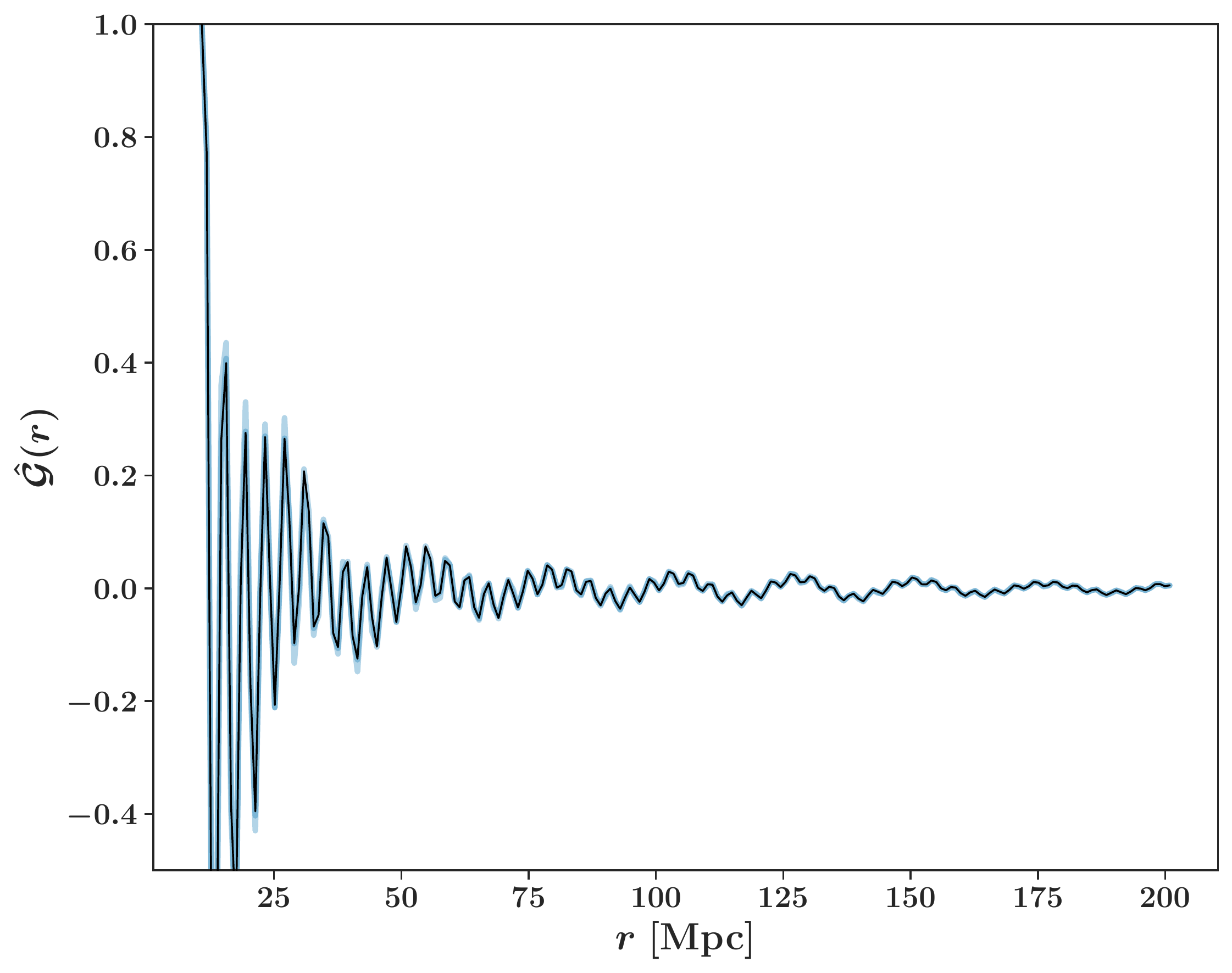}
    \caption{Green's function for the baryon-only transfer function in Figure~\ref{fig:transfer}. The shape depends on the value of the transfer function at all $k$. We extrapolate the transfer function from Figure~\ref{fig:transfer} using Gaussian process regression and the $\Lambda$CDM baryon transfer function as a prior. The black curve gives the mean result and the blue shows the standard deviation. The errors are dominated by the extrapolation choice, thus we exclude the statistical error bars.}
    \label{fig:greens}
\end{figure}

We show the associated Green's function, computed using the \texttt{hankel} python package\footnote{\url{https://github.com/steven-murray/hankel}}, in Figure~\ref{fig:greens}. Because the transform includes an integral over all $k$-modes, the exact form of the Green's function depends on the behavior of the transfer function outside of our data range. For the purposes of determining the Green's function, we need to extrapolate the high-$k$ range as it determines the small-$r$ behavior. We extrapolate the transfer function using Gaussian process regression from the \texttt{scikit-learn} python package\footnote{\url{https://scikit-learn.org/}} and assume the $\Lambda$CDM baryon transfer function, as computed with CAMB, as a prior.

Regardless of the extrapolation choice, the Green's function changes sign multiple times, including near the BAO scale. The Green's function shows the response a modified gravity theory of dark matter must have in order to explain the evolution of baryons on large scales. Thus, any alternative gravity theory would need to: 1) contain this scale to suppress the BAO features over time -- changing them from dominant at $z\sim 1100$ to very low amplitude at $z\sim 0.4$; and 2) have an acceleration law that changes sign around this scale.

\section{Conclusions \label{sec:conclusions}}
Cosmological observations place strong constraints on the form of any modification to General Relativity. In the absence of dark matter, the modified theory must explain how density fluctuations grow from the electron velocity field traced by the CMB polarization at $z=1100$ to the galaxy density field seen in the local universe. In this Letter, we show that any theory that depends linearly on the density field must have the peculiar Green's function shown in Figure \ref{fig:greens}. Given the extreme form of the function, it is not clear that it is possible to find such a theory -- in particular, the sign changes would induce quite extreme dynamics within the local volume \citep[for a recent work that performs this sort of analysis for Horndeski models, see][]{Rida2019}. While there are candidate modified gravity theories that fit the CMB temperature spectrum \citep{Skordis2020}, none have shown they can correctly predict the CMB polarization spectrum and the large-scale structure. CDM remains the simplest explanation for our cosmological observations.

\acknowledgements
The authors would like to thank the anonymous referee for their useful suggestions and feedback. We would also like to thank Miles Cranmer, Chen Heinrich, and Michael Strauss for helpful discussions and feedback on the manuscript. KP received support from the National Science Foundation Graduate Research Fellowship Program under grant DGE-1656466. This work was done as a private venture and not in KP's capacity as an employee of the Jet Propulsion Laboratory, California Institute of Technology.  Flatiron Institute is supported by the Simons Foundation.  We thank the authors of the following public software: Astropy \citep{astropy:2013, astropy:2018}, CAMB \citep{camb}, Hankel \citep{hankel}, Matplotlib \citep{Hunter:2007}, Numpy \citep{numpy}, Pandas \citep{pandas, mckinney-proc-scipy-2010}, Scikit-Learn \citep{scikit-learn}, SciPy \citep{scipy}, \& Seaborn \citep{seaborn}.  The software to reproduce the analysis and plots in this paper can be found at: \url{https://github.com/kpardo/mg_bao/}.

\bibliographystyle{apsrev4-2}
\bibliography{ref.bib}

\end{document}